\documentclass[sigconf,9pt,authorversion,nonacm]{acmart}

\AtBeginDocument{%
  \providecommand\BibTeX{{%
    \normalfont B\kern-0.5em{\scshape i\kern-0.25em b}\kern-0.8em\TeX}}}

\usepackage{tikz}
\newcommand\copyrighttext{%
  \footnotesize \textcopyright ACM, 2022. This is the author's version of the work. It is posted here by permission of ACM for your personal use. Not for redistribution. The definitive version was published in
  EdgeSys '22: Proceedings of the 5th International Workshop on Edge Systems, Analytics and Networking
  \href{https://dl.acm.org/doi/10.1145/3517206.3526275}{https://dl.acm.org/doi/10.1145/3517206.3526275}}
\newcommand\copyrightnotice{%
\begin{tikzpicture}[remember picture,overlay]
\node[anchor=south,yshift=10pt] at (current page.south) {\fbox{\parbox{\dimexpr\textwidth-\fboxsep-\fboxrule\relax}{\copyrighttext}}};
\end{tikzpicture}%
}

\copyrightyear{2022}
\acmYear{2022}
\setcopyright{acmlicensed}\acmConference[EdgeSys'22]{5th International Workshop on Edge Systems, Analytics and Networking }{April 5--8, 2022}{RENNES, France}
\acmBooktitle{5th International Workshop on Edge Systems, Analytics and Networking (EdgeSys'22), April 5--8, 2022, RENNES, France}
\acmPrice{15.00}
\acmDOI{10.1145/3517206.3526275}
\acmISBN{978-1-4503-9253-2/22/04}

\usepackage{hyperref}
\usepackage{siunitx}
\usepackage{enumitem}
\usepackage{framed}
\usepackage{subfig}
\usepackage{graphicx}

\newcounter{dummy}
\makeatletter
\newcommand\labeleditem[1][]{\item[#1]\refstepcounter{dummy}\def\@currentlabel{#1}}
\makeatother

\usepackage[nolist]{acronym}
\begin{acronym}
\acro{iot}[IoT]{Internet of Things}
\acro{faas}[FaaS]{Function-as-a-Service}
\acro{rest}[REST]{Representational state transfer}
\acro{paas}[PaaS]{Platform as a Service}
\acro{gcp}[GCP]{Google Cloud Platform}
\acro{cos}[COS]{Cloud Operating System}
\acro{RTT}[RTT]{Round Trip Time}
\acro{json}[JSON]{JavaScript Object Notation}
\acro{http}[HTTP]{Hyper Test Transfer Protocol}
\acro{ml}[ML]{Machine Learning}
\end{acronym}

\begin{document}

\title{SyncMesh: Improving Data Locality for Function-as-a-Service\\in Meshed Edge Networks}

\author{Daniel Habenicht}
\affiliation{%
  \institution{Technical University Berlin}
  \city{Berlin}
  \country{Germany}
}
\email{daniel.habenicht@campus.tu-berlin.de}

\author{Kevin Kreutz}
\affiliation{%
  \institution{Technical University Berlin}
  \city{Berlin}
  \country{Germany}
}
\email{kreutz@campus.tu-berlin.de}

\author{Soeren Becker}
\affiliation{%
  \institution{Technical University Berlin}
  \city{Berlin}
  \country{Germany}
}
\email{soeren.becker@tu-berlin.de}

\author{Jonathan Bader}
\affiliation{%
  \institution{Technical University Berlin}
  \city{Berlin}
  \country{Germany}
}
\email{jonathan.bader@tu-berlin.de}

\author{Lauritz Thamsen}
\affiliation{%
  \institution{University of Glasgow}
  \city{Glasgow}
  \country{United Kingdom}
}
\email{lauritz.thamsen@glasgow.ac.uk}

\author{Odej Kao}
\affiliation{%
  \institution{Technical University Berlin}
  \city{Berlin}
  \country{Germany}
}
\email{odej.kao@tu-berlin.de}

\begin{abstract}

The increasing use of \acl{iot} devices coincides with more communication and data movement in networks, which can exceed existing network capabilities.
These devices often process sensor or user information, where data privacy and latency are a major concern.
Therefore, traditional approaches like cloud computing do not fit well, yet new architectures such as edge computing address this gap. In addition, the \ac{faas} paradigm gains in prevalence as a workload execution platform, however the decoupling of storage results in further challenges for highly distributed edge environments.

To address this, we propose SyncMesh, a system to manage, query, and transform data in a scalable and stateless manner by leveraging the capabilities of \acl{faas} and at the same time enabling data locality. 
Furthermore, we provide a prototypical implementation and evaluate it against established centralized and decentralized systems in regard to traffic usage and request times. 

The preliminary results indicate that SyncMesh is able to exonerate the network layer and accelerate the transmission of data to clients, while simultaneously improving local data processing.
\end{abstract}



\maketitle
\copyrightnotice

\section{Introduction}
\label{sec:introduction}
With the increase in computational power and popularity of smart mobile devices, the \ac{iot} has been gaining momentum in the past years. In turn, aspects like scalability \cite{gupta2017scalability} and locality of data \cite{vaquero2014finding} are becoming more challenging \cite{ChiangOverviewFogIoT}. 
Therefore, cloud infrastructures are getting increasingly distributed to support localized edge and fog architectures that manage requests in close proximity to the users and data~\cite{shi2016promise}. 
These architectures reduce response latencies and at the same time unburden the network layer \cite{IFCIoTintegratedFogCloudIoT,DitasApproach}. 
In order to cope with the increasing scale and the dynamic nature of \ac{iot} environments, ongoing research~\cite{meshnetworks, liu2017wireless} proposes to integrate ad-hoc mesh networks and decentralized capabilities to further enhance the scalability and resilience of edge and \ac{iot} architectures. 
In addition, the Serverless and \ac{faas} paradigm has been identified as a suitable computing model for the \ac{iot} \cite{nastic2017serverless}, since it simplifies the deployment at the edge, enables on-demand allocation of resources, and allows for the execution of tasks in lightweight containers~\cite{tinyFaas}. 

In the \ac{faas} paradigm, computation and storage are typically decoupled: The functions themselves are stateless and the used data or results are stored in distributed storage systems. Although distributed file systems (DFS) or object stores like S3 simplify the usage of \ac{faas} in cloud computing environments, the dynamic nature of upcoming architectures in the edge-cloud continuum poses new challenges: Devices are often connected via 
an unreliable and possibly slow network connection, only provided with limited hardware resources, and can join and leave the network at any time in an ad-hoc manner \cite{ferrerCognitiveComputeContinuum2021a}. 

Summarizing, Serverless and \ac{faas} enable a scalable workload execution, but rely on external storage to store and retrieve state, typically outsourced to centralized cloud systems \cite{eismann2020serverless}. The possible impact of the external storage for \ac{faas} and Serverless has been identified as a major issue in cloud architectures \cite{oneStepFowardTwoStepsBack}, yet is even further exacerbated in dynamic \ac{iot} and edge environments.

In an effort to address this challenge, this paper presents \emph{SyncMesh}, a new system that improves data locality in mesh-based edge and fog computing environments for \ac{faas} workloads. Instead of moving all generated data of edge devices, i.e. sensor readings, to a central cloud storage, the information is stored on the respective nodes, analyzed, aggregated and transformed locally, and provided to other nodes in the network on-demand. Our prototype implementation of SyncMesh offers a \ac{faas} interface for local data processing and is able to exchange data between participating nodes in a sensor network, significantly reducing network transmissions in the highly distributed and dynamic edge/fog environments.

In summary, the main contributions of this paper are:
\begin{itemize}
    \item We outline assumptions, challenges and requirements of data storage and locality in highly distributed fog and edge environments.
    \item We derive the \emph{SyncMesh} architecture and implement a prototype\footnote{\url{https://github.com/dos-group/SyncMesh}} to tackle the identified challenges and enable data locality as well as in-situ data processing.
    \item We present a preliminary evaluation of our prototype by implementing a relevant use case using real world data and comparing our results to different baseline scenarios.
\end{itemize}

The remainder of this paper is structured as followed:
We first provide an overview of the related work before stating our assumptions and discussing the requirements. Subsequently, we present the \emph{SyncMesh} prototype, conducted experiments as well as results before concluding the paper.

\section{Related Work}
\label{sec:related-work}
In general, the importance of storage and data management for
the upcoming distributed environments is growing, not only due to the ever increasing generated data \cite{Moysiadis2018TowardsComputing}. Subsequently, several approaches to improve the storage layer in fog and edge computing environments exists:
Many systems focus on efficient cloud offloading~\cite{bermbach2021auctionwhisk, Wang2019Fog-basedIoT}, in which advanced algorithms are used to time the offloading of data to the cloud efficiently \cite{SheikhSofla2021TowardsComputing}. In contrast to this, with \emph{SyncMesh} our goal is to store and analyze the data where it is created. 

In \cite{confais1}, the authors evaluate the performance of different object store system in edge and fog environments, while also identifying storage as a major challenge. In subsequent works, Confais et. al.\cite{confais2} propose a combination of a network attached storage and peer-to-peer technology to improve the performance of object stores in the aforementioned environments. Although presenting promising results, their approach still relies on a distributed storage layer available to the respective edge nodes, whereas \emph{SyncMesh} aims to improve the data locality.

More similar to our approach, other works avoid the use of central instances entirely and rely on P2P communication between the devices~\cite{p2pEdgeDataLocality}. Mayer et al. \cite{fogstore} address data management problems within decentralized fog networks and introduce their own solution called FogStore, which is a context-aware distributed data storage system. In addition to state management, the paper outlines specific replica placement strategies and a generalized API for querying and manipulating data. In contrast, in our approach we neglect replicas and rather only exchange data between nodes if needed.

Furthermore, the Nebulastream platform \cite{zeuch2019nebulastream} was designed with the specific purpose of usage as a data management system for the \ac{iot} and addresses issues arising from centralized approaches. They rely on stream processing instead of \ac{faas}.

In regard to \ac{faas}, the Fog function project \cite{fogfunction} explores \ac{faas} as a serverless programming model within the context of \ac{iot} and uses event-listeners to discover and orchestrate devices and resources in \ac{iot} environments. In addition, Fog function was integrated into FogFlow, which is a framework for fog computing with the goal of providing a programming model that enables programmers to develop \ac{iot} systems more easily~\cite{Cheng2018FogFlow:Cities}.

Although the \ac{faas} paradigm is often highlighted as fitting workload and data analytics platform for fog and edge computing \cite{nastic2017serverless}, most are focused on i.e. enabling a lightweight orchestration \cite{tinyFaas} or scheduling based on resource availabilities \cite{rausch2021optimized} and often not take the location of data into account. With \emph{SyncMesh} we attempt to address this challenge, to further improve the data locality.

\section{SyncMesh}
\label{sec:SyncMesh}

In this section, we present the design of the \emph{SyncMesh} system. Firstly, we describe assumptions about the computing environment before we derive a set of requirements for our system. Subsequently, the different components of the \emph{SyncMesh} system are introduced.
\subsection{Assumptions}
\label{sec:assumptions}
In order to define the scope and envisioned use case of \emph{SyncMesh}, we make the following assumptions about the expected environment:
We assume a highly distributed environment in the edge-cloud continuum, consisting of lightweight and heterogeneous devices. The devices are interconnected via peer-to-peer connections in an ad-hoc mesh network and act as autonomous nodes. Therefore, the environment is not administered or managed by a central entity, and devices organize themselves in, i.e. swarms, in order to enable scenarios in the context of remote sensing and environmental awareness \cite{ferrerCognitiveComputeContinuum2021a}. In the related work, such architectures are often proposed to increase the scalability and reliability of fog and edge computing environments and to cope with the dynamic nature of the \ac{iot}. 
 
Furthermore, we assume that the devices are equipped with sensors such as temperature and air quality sensors or audio and video sources, resulting in data directly created on the respective nodes.
The devices are not constrained to data center boundaries and can be located anywhere in i.e. a smart city or a wildlife refuge \cite{ayele2018towards}, and due to the possibly unreliable and slow network it is not feasible to constantly synchronize data between nodes or upload all sensor data to the cloud.

Finally, for our initial prototype we presume that inside the network of edge devices, a discovery process for neighboring nodes in the mesh exists. Therefore, for the remainder of this paper we expect that devices can join the network and automatically discover other nodes in the proximity, but neglect an actual implementation and refer
to the related work. For instance, previous work \cite{becker2021local} has utilized metrics provided by the underlying mesh network to identify direct neighbors in the ad-hoc swarm.

\subsection{Requirements}
\label{sec:SyncMesh:requirements}
From our assumptions, we derive the following requirements for our system:

\paragraph{Support operation during unreliable network conditions}
The described edge sites are often distributed across several remote locations and are only connected with limited bandwidth and high latencies. In addition, network partitions and outages are more common than in traditional data centers. Therefore, the system needs to be able to cope with and automatically adjust to poor network connectivity without intrusive impact on the local processing of data.

\paragraph{Enable local data processing and storage} 
In order to not overload the network layer and subsequently interfere with other and possibly critical network traffic, the data needs to be stored, processed, transformed, or aggregated where it is created. Thus, data locality needs to be enabled and the local processing simplified.

\paragraph{Autonomous operation and resilience} 
Due to the dynamic nature of the aforementioned \ac{iot} environments, lightweight edge nodes are expected to join and leave the network at any time. Therefore, the system needs to cope automatically with node joins, churns, and failures by, i.e., enabling the entities to work as autonomous as possible.

\paragraph{Environmental awareness} 
Although the devices are expected to work in an autonomous manner, to increase the remote sensing capabilities and subsequently the environmental awareness, nodes need to be able to exchange their raw or aggregated data on-demand with other nodes in the proximity.

\subsection{System Architecture}
\label{sec:SyncMesh:prototype}

With the given assumptions on the environment and subsequently to fit the requirements stated before, we implemented a prototype based on lightweight containerization technologies, which
improves the data locality and simplifies the deployment or workload by utilizing \ac{faas} capabilities.

Therefore, each device in the environment -- in the following called a \emph{SyncMesh node} -- acts as an autonomous node in the mesh network, stores sensor data, and offers an interface that enables users and other nodes in the proximity to collaborate with it.

\begin{figure}[!h]
	\centering
		\includegraphics[width=\linewidth]{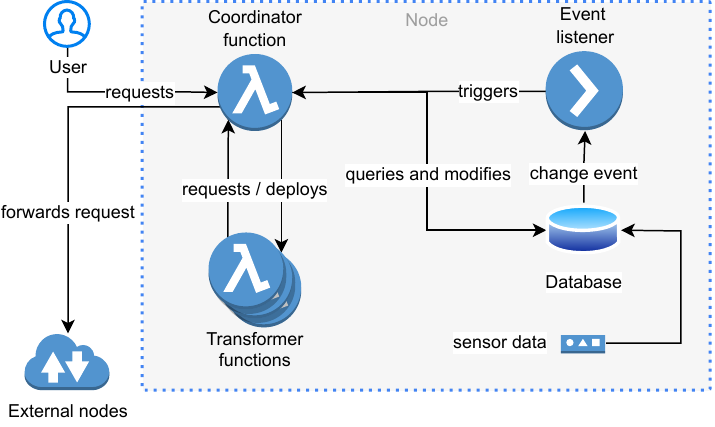}
	\caption{Architecture and key components of a SyncMesh node.}
    \label{fig:SyncMesh-node}
\end{figure}

A \emph{SyncMesh node} contains several key components, which will be discussed in more detail in the remainder of this section.

\paragraph{Local database}
Each node employs a local database which is used to store data generated by connected sensors and the availability model of other available devices in the proximity. The database can further be accessed by the coordinator function and other available functions on the device and offers the possibility to use event listeners to, i.e., automatically react to changes in the database. In our prototype, we use a MongoDB instance, although other solutions can be used where appropriate.

\paragraph{Database event listener}
A separately instantiated database listener registers new sensor data input and forwards the change events to the coordinator function. Therefore, it provides notification functionalities, sharing database changes to subscribed external nodes or calling (transformer) functions for additional processing.

\paragraph{Coordinator function}
The \emph{SyncMesh} interface is exposed via the coordinator function
and is used as an entry point for users and other nodes in the sensor network to access data on the respective node. In general, it is a serverless function implemented in the OpenFaaS framework and provided via an HTTP API.
Every request uses the same unified request schema containing the data query as a GraphQL statement, as well as additional parameters like for instance transformer functions.

The coordinator function can store and modify data on the node itself by interacting with the local database.
In addition, it forwards events registered by the database listener to subscribed nodes, enabling decentralized data replication as well as event-driven mechanisms. 
Although the coordinator function already implements aggregation functionalities, it forwards more advanced
data pre-processing tasks to other local transformer functions that enable diverse workloads, from simple analysis to \ac{ml} workloads.
Finally, before sending the data to the client, the response is compressed with gzip.

\paragraph{Transformer functions}
In order to reduce the data transmission between nodes and to
facilitate operation in limited bandwidth scenarios, transformer functions can be used to pre-process data directly on the respective
nodes. Sensor data can, i.e., be aggregated and analyzed with \ac{ml} models, e.g., image classifiers or other data analysis pipelines, before being transferred over the network.

The envisioned workload is implemented in a function or a chain of functions, provided in docker containers and 
started by the coordinator function in case of changes in the database or when triggered externally, e.g. by nodes in the proximity. 
The respective containers are scaled on-demand as well as stopped in case they are not needed anymore, therefore relieving resources of lightweight edge nodes.

\paragraph{Requirement analysis}
In summary, in order to support the operation during unreliable network conditions, \emph{SyncMesh} nodes only exchange data on-demand and favor local data pre-processing as much as possible, subsequently relieving the network layer. 
The sensor data is initially only stored on the respective devices, and by leveraging \ac{faas} capabilities, the local processing is simplified via the deployment of arbitrary (transformer) functions. Due to the stateless nature of the serverless functions, they can be scaled up and down if needed, thus automatically adapting to increasing and decreasing demand. Therefore, a single \emph{SyncMesh} node acts autonomously in regard to processing the locally available (sensor) data.

Further, the data can also be exchanged and replicated across other available nodes in the environment on-demand. 
For that reason, each \emph{SyncMesh} node maintains a model of other available devices in the proximity \cite{becker2021local} and transparently forwards user requests regarding sensor readings in the surrounding area to respective neighboring nodes, possibly aggregating the results and returning a unified response. 
The requests are only forwarded to currently available nodes.
Consequently, \emph{SyncMesh} is automatically adjusting to node churns and joins.

Finally, since all components and used functions in \emph{Syncmesh} are provided as containers, they can be adapted to heterogeneous CPU architectures by i.e. providing multi-arch Docker images \footnote{https://docs.docker.com/desktop/multi-arch/}.

\section{Evaluation}
\label{sec:evaluation}
In order to evaluate our \emph{SyncMesh} prototype, we conducted a set of experiments, comparing it to several baselines and describe our results in this section.

\subsection{Experiment setup}
\begin{figure}[!h]
	\centering
		\includegraphics[width=0.5\linewidth]{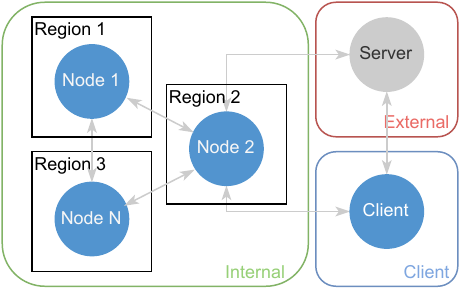}
	\caption{A basic network diagram visualizing the experiment setup.}
    \label{fig:experiment}
\end{figure}

To simulate an environment as described in our assumptions, we created a virtual testbed on the Google Cloud Platform. The hardware and software specifications for each used virtual machine instance are listed in \autoref{tbl:specification}. As depicted in \autoref{fig:experiment}, we deployed virtual machines in different
cloud regions to introduce latencies between participating nodes in a distributed environment. Subsequently this resulted in latencies ranging between 20 and 300ms. In addition, in some cases we used a separate virtual machine as an external server for the implemented baselines.

In order to simulate sensors connected to the edge devices, we leveraged a real-world \ac{iot} dataset, containing air quality, temperature and humidity measurements collected in a city in Bulgaria \footnote{\url{https://airsofia.info/}}. The dataset was distributed across the nodes and added to the respective databases, simulating real world measurements.

For the evaluation, we considered two different scenarios: 
\begin{itemize}
    \item Collect: A client requests all data from all nodes of a sensor network for a given time span. Therefore, the sensor data is retrieved from all participating nodes and provided to the user.
    \item Transform: A client requests i.e. aggregated data from the surroundings of an edge node. Therefore, the sensor data is, if applicable, preprocessed and then provided to the user.
\end{itemize}

As illustrated in \autoref{fig:experiment}, we monitored the used traffic for requests from the client (blue), internal network traffic between the nodes (green), and in case of baselines from and to the external cloud server (red). In addition, we used the \ac{RTT} until a clients request was successfully processed as a further evaluation metric.

Finally, we performed the experiments for different amounts of nodes in the network (3, 6, 9 and 12) and several data time spans (1, 7, 14 and 30 days). 
Each configuration was repeated 20 times to improve accuracy and soften outliers.

The used scripts, software and additional tooling are provided in our GitHub repository. %

\begin{table}[!h]
\caption{Hard and Software Specifications}
\centering
\begin{tabular}{|c|c|}
   \hline
   \multicolumn{2}{|c|}{Hardware (each GCP VM)} \\ \hline
  CPU & 1vCPU @ 2.60GHz \\
   Memory & 3.75 GB \\ 
   Storage & 20 GB\\\hline
    \multicolumn{2}{|c|}{Software} \\ \hline
    Ubuntu & \href{https://cloud.google.com/compute/docs/images/os-details#ubuntu_lts}{20.04-lts} \\ 
    MongoDB & 5.0.2 \\
    GUN & 0.2020.1235 \\
    Node & 14 \\ 
    faasd & 0.14.2 \\ \hline
\end{tabular}
\label{tbl:specification}
\end{table}

\subsection{Baselines}
\label{sec:evaluation:systems}
We implemented three other systems to compare them against \emph{SyncMesh}.
These baselines were chosen because they are being used in practice to handle vast amounts of data in \ac{iot} networks. %

\begin{figure*}[!t]
\centering
\subfloat[Time until all sensor data of a three node cluster is retrieved by the client, including the results of the p2p baseline. The y-axis is log scaled to enable the representation in a single plot.]{
\includegraphics[width=0.3\textwidth]{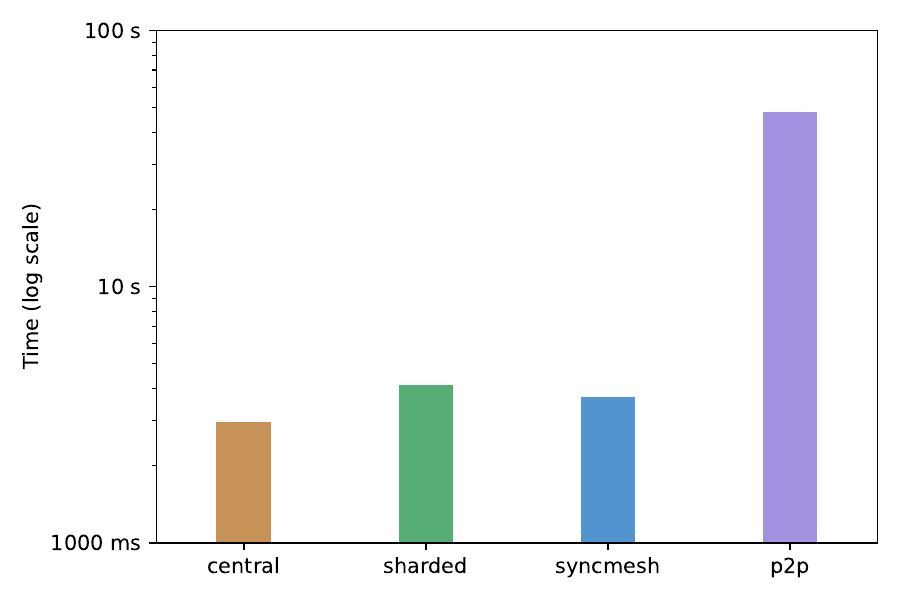}
\label{fig:request_time:overview}
}\hfil   
\subfloat[Distribution of network traffic for the collect scenario on different systems in a sensor network consisting of three nodes. 
]{
\includegraphics[width=0.3\textwidth]{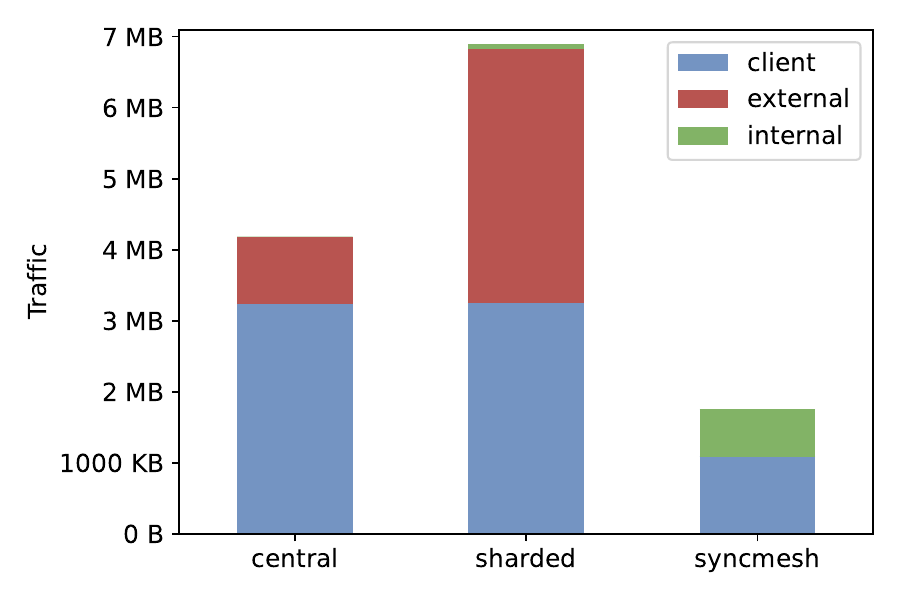}
\label{fig:traffic:collect}
}\hfil
\subfloat[Distribution of network traffic for the transform scenario (average sensor readings) in a sensor network consisting of three nodes.
]{
\includegraphics[width=0.3\textwidth]{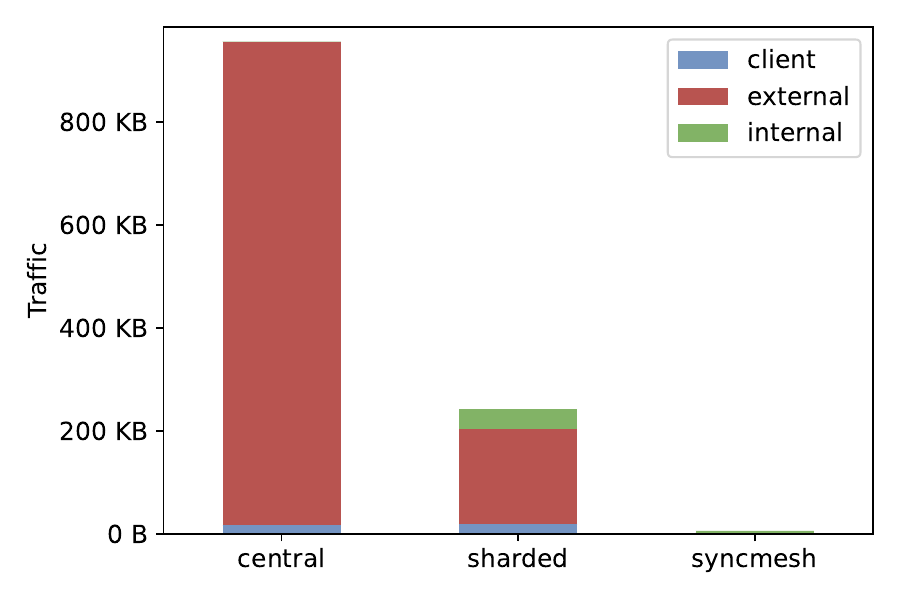}
\label{fig:traffic:transform}
}
\caption{
Exemplary request time for a three node network including the p2p baseline (a) and the distribution of monitored network traffic between nodes in the network, the client and respectively a central server during the collect (b) and transform (c) scenario.
}
\label{fig:traffic}
\end{figure*}

\begin{figure*}%

\centering
\subfloat[Time needed to retrieve all data from a sensor network, for different baselines, network sizes and the collect scenario (30 days). For the baselines, the graph also includes the time needed to send the data to i.e. the central database.]{
\includegraphics[width=0.3\textwidth]{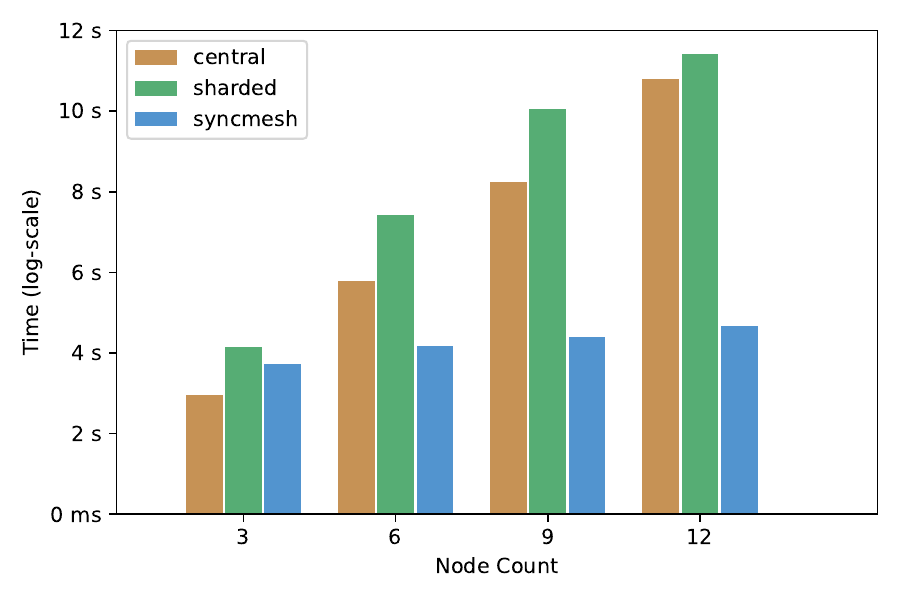}
\label{fig:compare:rtt:node}
}\hfil
\subfloat[Used traffic in the sensor network for different baselines, network sizes and the collect scenario (30 days).]{
\includegraphics[width=0.3\textwidth]{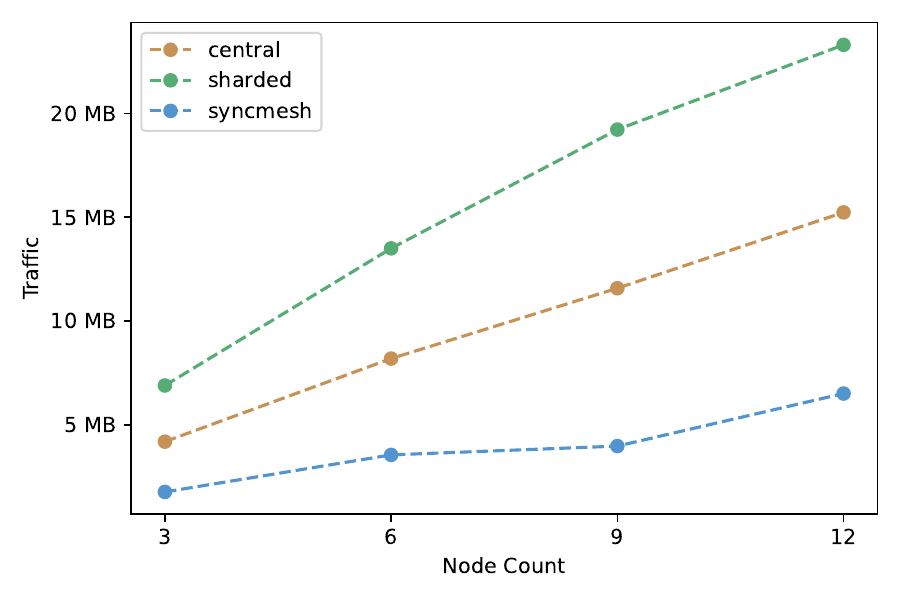}
\label{fig:compare:node:collect}
}\hfil 
\subfloat[Used traffic in the sensor network for different baselines, network sizes and the transform scenario (average sensor readings of last 30 days).]{
\includegraphics[width=0.3\textwidth]{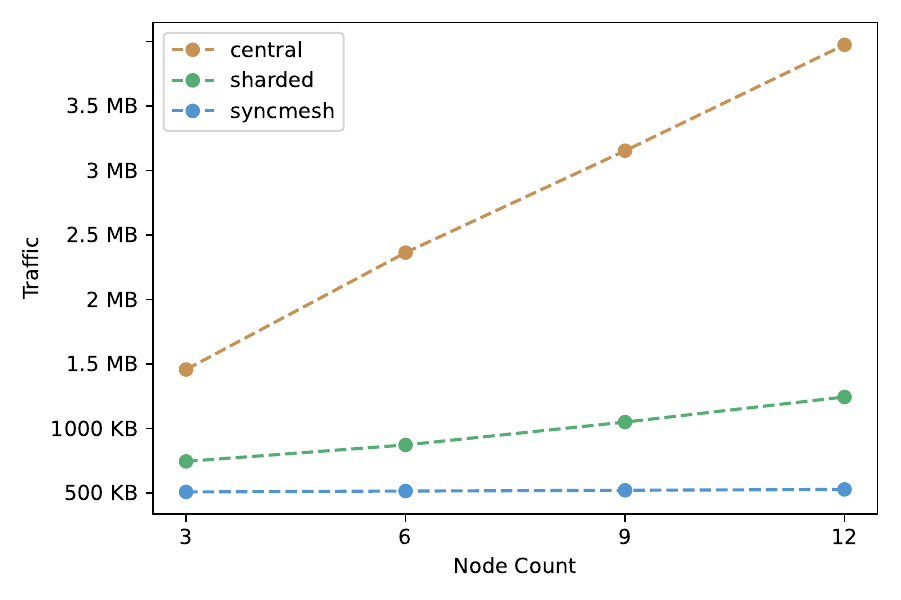}
\label{fig:compare:node:aggregate}
} 
\caption{Request time (a) and combined traffic between nodes, the client and respectively external server for the collect (b) and transform (c) scenarios in different network sizes.} 
\label{figure:compare}
\end{figure*}

\begin{description}
    \labeleditem[central]cloud storage \label{system:central} \\
    A standard implementation of a sensor network. Each sensor sends the data to a central cloud database (MongoDB) from where clients can retrieve it.
    \labeleditem[sharded]cloud storage \label{system:sharded} \\
    A sharded implementation of the central cloud storage. Each sensor stores its own data, and a central server is supplying unified access to the client. Each sensor thereby represents a shard in the database. 
    \labeleditem[p2p]distributed database storage \label{system:peer2peer} \\
    A peer-to-peer database with eventual consistency based on GunDB\footnote{https://gun.eco/}. 
    Clients need to individually connect to peers in the network in order to retrieve all sensor data or synchronize data between nodes.
\end{description}

In contrast to the other baselines, \ref{system:peer2peer} sends all data uncompressed. 
The MongoDB based systems use snappy and \emph{SyncMesh} implements a gzip compression.

\subsection{Results}
In this section we first discuss the measurements of our comparison benchmark.

\autoref{fig:traffic} describes the evaluation results of a three node sensor network. 
Therefore, \autoref{fig:request_time:overview} depicts the time a client needs to retrieve all sensor data from all nodes, when using different baselines and \emph{SyncMesh}.
As can be seen, the \emph{p2p} baseline needs significant more time than all other baselines, which is why we excluded it in other plots to improve the detectability. 
Furthermore, \emph{SyncMesh} performs 10.1\% better than \emph{sharded} and 20.7\% worse than the \emph{central}
baseline.

In \autoref{fig:traffic:collect} we show the produced traffic for each systems during the collect scenario, except for the \emph{p2p} baseline (91.03 MB). The results further show that \emph{SyncMesh} is able to outperform the \emph{sharded} and \emph{central} baselines by up to 57.8\% in regard to network efficiency. This is in part due to the different compression algorithms used in MongoDB and \emph{SyncMesh}. In the case of \emph{p2p} the amount of data sent is even doubled, due to protocol specifics of the employed solution. In addition, \emph{SyncMesh} therefore also produces less internal traffic between participating nodes than all of the other solutions. 

This is especially evident in \autoref{fig:traffic:transform}, which depicts the used traffic during the transform scenario.
\emph{SyncMesh} is again able to outperform the baselines by only using \SI{6.99} kB. As can be seen, the majority of used traffic results from transmitting the sensor data to the central server.
In contrast to \autoref{fig:traffic:transform}, the \ref{system:sharded} system performs 75.0\% better than the \ref{system:central} system, as the sharded MongoDB also uses local workers (for aggregation) on each shard, decreasing the amount of traffic between the server and its shards. 

The plots in \autoref{figure:compare} describe the same experiment as \autoref{fig:traffic}, but for a three, six, nine and twelve node network respectively.
As expected and depicted in \autoref{fig:request_time:overview}, the \emph{sharded} and \emph{central} baseline show a continuous increase in request times for growing network sizes due to the increased data amount that must be send to the cloud from all nodes and retrieved by the client. 
However, the performance of \emph{SyncMesh} does not degrade significantly: The time is only increased by 25.7\%, in comparison to 175.6\% for \emph{shared} and  266.1\% in case of \emph{central}, most likely due to the decreased traffic between nodes. 

This trend is also observable in \autoref{fig:compare:node:collect} and \autoref{fig:compare:node:aggregate}, that describe the used traffic for the collect and transform scenarios across different network sizes. As can be seen, \emph{SyncMesh} scales significantly better in regard to transmitted traffic, and therefore is able to exonerate the network link between participating nodes in the network.

\section{Conclusion}

In this paper we have introduced \emph{SyncMesh}, a system that improves the data locality and enables local data processing based on the \ac{faas} paradigm in a meshed network of edge nodes. We prototypical implemented \emph{SyncMesh} and evaluated our approach in a virtualized testbed on a real world \ac{iot} dataset, and compared it against several  baselines in regard to request time and network utilization.

As indicated by the preliminary results, \emph{SyncMesh} is
able to outperform the traditional approaches in regard to network usage and request times, in particular for larger amounts of nodes. 

Consequently, our results underline the importance and impact of data locality, especially for the the application of \ac{faas} in highly distributed environments.

In future work we plan to further extend the \emph{SyncMesh} prototype, include more complex analysis workloads and evaluate our approach on real-world edge devices. In addition, we intend to assess the impact of cold starts in the \ac{faas} framework.

\begin{acks}
We would like to thank Denis Koljada for contributing to the initial prototype within the scope of a master’s project.
\end{acks}

\bibliographystyle{ACM-Reference-Format}
\bibliography{mendeley}


\begin{thebibliography}{27}


\ifx \showCODEN    \undefined \def \showCODEN     #1{\unskip}     \fi
\ifx \showDOI      \undefined \def \showDOI       #1{#1}\fi
\ifx \showISBNx    \undefined \def \showISBNx     #1{\unskip}     \fi
\ifx \showISBNxiii \undefined \def \showISBNxiii  #1{\unskip}     \fi
\ifx \showISSN     \undefined \def \showISSN      #1{\unskip}     \fi
\ifx \showLCCN     \undefined \def \showLCCN      #1{\unskip}     \fi
\ifx \shownote     \undefined \def \shownote      #1{#1}          \fi
\ifx \showarticletitle \undefined \def \showarticletitle #1{#1}   \fi
\ifx \showURL      \undefined \def \showURL       {\relax}        \fi
\providecommand\bibfield[2]{#2}
\providecommand\bibinfo[2]{#2}
\providecommand\natexlab[1]{#1}
\providecommand\showeprint[2][]{arXiv:#2}

\bibitem[\protect\citeauthoryear{Ayele, Meratnia, and Havinga}{Ayele
  et~al\mbox{.}}{2018}]%
        {ayele2018towards}
\bibfield{author}{\bibinfo{person}{Eyuel~D Ayele}, \bibinfo{person}{Nirvana
  Meratnia}, {and} \bibinfo{person}{Paul~JM Havinga}.}
  \bibinfo{year}{2018}\natexlab{}.
\newblock \showarticletitle{Towards a new opportunistic iot network
  architecture for wildlife monitoring system}. In
  \bibinfo{booktitle}{\emph{NTMS}}. \bibinfo{publisher}{IEEE}.
\newblock


\bibitem[\protect\citeauthoryear{Becker, Schmidt, Thamsen, Ferrer, and
  Kao}{Becker et~al\mbox{.}}{2021}]%
        {becker2021local}
\bibfield{author}{\bibinfo{person}{Soeren Becker}, \bibinfo{person}{Florian
  Schmidt}, \bibinfo{person}{Lauritz Thamsen}, \bibinfo{person}{Ana~Juan
  Ferrer}, {and} \bibinfo{person}{Odej Kao}.} \bibinfo{year}{2021}\natexlab{}.
\newblock \showarticletitle{{{LOS}}: {{Local-Optimistic Scheduling}} of
  {{Periodic Model Training For Anomaly Detection}} on {{Sensor Data Streams}}
  in {{Meshed Edge Networks}}}. In \bibinfo{booktitle}{\emph{ACSOS}}.
  \bibinfo{publisher}{{IEEE}}.
\newblock


\bibitem[\protect\citeauthoryear{Bermbach, Bader, Hasenburg, Pfandzelter, and
  Thamsen}{Bermbach et~al\mbox{.}}{2021}]%
        {bermbach2021auctionwhisk}
\bibfield{author}{\bibinfo{person}{David Bermbach}, \bibinfo{person}{Jonathan
  Bader}, \bibinfo{person}{Jonathan Hasenburg}, \bibinfo{person}{Tobias
  Pfandzelter}, {and} \bibinfo{person}{Lauritz Thamsen}.}
  \bibinfo{year}{2021}\natexlab{}.
\newblock \showarticletitle{AuctionWhisk: Using an auction-inspired approach
  for function placement in serverless fog platforms}.
\newblock \bibinfo{journal}{\emph{Software: Practice and Experience}}
  (\bibinfo{year}{2021}).
\newblock


\bibitem[\protect\citeauthoryear{Burchard, Chemodanov, Gillis, and
  Calyam}{Burchard et~al\mbox{.}}{2017}]%
        {meshnetworks}
\bibfield{author}{\bibinfo{person}{Josiah Burchard}, \bibinfo{person}{Dmitrii
  Chemodanov}, \bibinfo{person}{John Gillis}, {and} \bibinfo{person}{Prasad
  Calyam}.} \bibinfo{year}{2017}\natexlab{}.
\newblock \showarticletitle{Wireless Mesh networking Protocol for sustained
  throughput in edge computing}. In \bibinfo{booktitle}{\emph{ICNC}}.
  \bibinfo{publisher}{IEEE}.
\newblock


\bibitem[\protect\citeauthoryear{Cheng, Fuerst, Solmaz, and Sanada}{Cheng
  et~al\mbox{.}}{2019}]%
        {fogfunction}
\bibfield{author}{\bibinfo{person}{Bin Cheng}, \bibinfo{person}{Jonathan
  Fuerst}, \bibinfo{person}{Gurkan Solmaz}, {and} \bibinfo{person}{Takuya
  Sanada}.} \bibinfo{year}{2019}\natexlab{}.
\newblock \showarticletitle{{Fog function: Serverless fog computing for data
  intensive IoT services}}. In \bibinfo{booktitle}{\emph{SCC}}.
  \bibinfo{publisher}{IEEE}.
\newblock


\bibitem[\protect\citeauthoryear{Cheng, Solmaz, Cirillo, Kovacs, Terasawa, and
  Kitazawa}{Cheng et~al\mbox{.}}{2018}]%
        {Cheng2018FogFlow:Cities}
\bibfield{author}{\bibinfo{person}{Bin Cheng}, \bibinfo{person}{Gürkan
  Solmaz}, \bibinfo{person}{Flavio Cirillo}, \bibinfo{person}{Ernö Kovacs},
  \bibinfo{person}{Kazuyuki Terasawa}, {and} \bibinfo{person}{Atsushi
  Kitazawa}.} \bibinfo{year}{2018}\natexlab{}.
\newblock \showarticletitle{{FogFlow: Easy Programming of IoT Services Over
  Cloud and Edges for Smart Cities}}.
\newblock \bibinfo{journal}{\emph{IEEE Internet of Things Journal}}
  \bibinfo{volume}{5}, \bibinfo{number}{2} (\bibinfo{year}{2018}).
\newblock


\bibitem[\protect\citeauthoryear{Chiang and Zhang}{Chiang and Zhang}{2016}]%
        {ChiangOverviewFogIoT}
\bibfield{author}{\bibinfo{person}{Mung Chiang} {and} \bibinfo{person}{Tao
  Zhang}.} \bibinfo{year}{2016}\natexlab{}.
\newblock \showarticletitle{{Fog and IoT: An Overview of Research
  Opportunities}}.
\newblock \bibinfo{journal}{\emph{IEEE Internet of Things Journal}}
  \bibinfo{volume}{3}, \bibinfo{number}{6} (\bibinfo{year}{2016}).
\newblock


\bibitem[\protect\citeauthoryear{Confais, Lebre, and Parrein}{Confais
  et~al\mbox{.}}{[n.\,d.]}]%
        {confais2}
\bibfield{author}{\bibinfo{person}{Bastien Confais}, \bibinfo{person}{Adrien
  Lebre}, {and} \bibinfo{person}{Benoît Parrein}.}
  \bibinfo{year}{[n.\,d.]}\natexlab{}.
\newblock \showarticletitle{An Object Store Service for a Fog/Edge Computing
  Infrastructure Based on IPFS and a Scale-Out NAS}. In
  \bibinfo{booktitle}{\emph{ICFEC}}. \bibinfo{publisher}{IEEE}.
\newblock


\bibitem[\protect\citeauthoryear{Confais, Lebre, and Parrein}{Confais
  et~al\mbox{.}}{2016}]%
        {confais1}
\bibfield{author}{\bibinfo{person}{Bastien Confais}, \bibinfo{person}{Adrien
  Lebre}, {and} \bibinfo{person}{Benoît Parrein}.}
  \bibinfo{year}{2016}\natexlab{}.
\newblock \showarticletitle{Performance Analysis of Object Store Systems in a
  Fog/Edge Computing Infrastructures}. In \bibinfo{booktitle}{\emph{CloudCom}}.
  \bibinfo{publisher}{IEEE}.
\newblock


\bibitem[\protect\citeauthoryear{Eismann, Scheuner, Van~Eyk, Schwinger,
  Grohmann, Herbst, Abad, and Iosup}{Eismann et~al\mbox{.}}{2020}]%
        {eismann2020serverless}
\bibfield{author}{\bibinfo{person}{Simon Eismann}, \bibinfo{person}{Joel
  Scheuner}, \bibinfo{person}{Erwin Van~Eyk}, \bibinfo{person}{Maximilian
  Schwinger}, \bibinfo{person}{Johannes Grohmann}, \bibinfo{person}{Nikolas
  Herbst}, \bibinfo{person}{Cristina~L Abad}, {and} \bibinfo{person}{Alexandru
  Iosup}.} \bibinfo{year}{2020}\natexlab{}.
\newblock \showarticletitle{Serverless applications: Why, when, and how?}
\newblock \bibinfo{journal}{\emph{IEEE Software}} \bibinfo{volume}{38},
  \bibinfo{number}{1} (\bibinfo{year}{2020}).
\newblock


\bibitem[\protect\citeauthoryear{Ferrer, Becker, Schmidt, Thamsen, and
  Kao}{Ferrer et~al\mbox{.}}{2021}]%
        {ferrerCognitiveComputeContinuum2021a}
\bibfield{author}{\bibinfo{person}{Ana~Juan Ferrer}, \bibinfo{person}{Soren
  Becker}, \bibinfo{person}{Florian Schmidt}, \bibinfo{person}{Lauritz
  Thamsen}, {and} \bibinfo{person}{Odej Kao}.} \bibinfo{year}{2021}\natexlab{}.
\newblock \showarticletitle{Towards a {{Cognitive Compute Continuum}}: {{An
  Architecture}} for {{Ad-Hoc Self-Managed Swarms}}}. In
  \bibinfo{booktitle}{\emph{CCGrid}}. \bibinfo{publisher}{{IEEE}}.
\newblock


\bibitem[\protect\citeauthoryear{Gupta, Christie, and Manjula}{Gupta
  et~al\mbox{.}}{2017}]%
        {gupta2017scalability}
\bibfield{author}{\bibinfo{person}{Anisha Gupta}, \bibinfo{person}{Rivana
  Christie}, {and} \bibinfo{person}{PR Manjula}.}
  \bibinfo{year}{2017}\natexlab{}.
\newblock \showarticletitle{Scalability in internet of things: features,
  techniques and research challenges}.
\newblock \bibinfo{journal}{\emph{Int. J. Comput. Intell. Res}}
  \bibinfo{volume}{13}, \bibinfo{number}{7} (\bibinfo{year}{2017}).
\newblock


\bibitem[\protect\citeauthoryear{Hellerstein, Faleiro, Gonzalez,
  Schleier-Smith, Sreekanti, Tumanov, and Wu}{Hellerstein
  et~al\mbox{.}}{2019}]%
        {oneStepFowardTwoStepsBack}
\bibfield{author}{\bibinfo{person}{Joseph~M. Hellerstein},
  \bibinfo{person}{Jose Faleiro}, \bibinfo{person}{Joseph~E. Gonzalez},
  \bibinfo{person}{Johann Schleier-Smith}, \bibinfo{person}{Vikram Sreekanti},
  \bibinfo{person}{Alexey Tumanov}, {and} \bibinfo{person}{Chenggang Wu}.}
  \bibinfo{year}{2019}\natexlab{}.
\newblock \showarticletitle{{Serverless computing: One step forward, two steps
  back}}. In \bibinfo{booktitle}{\emph{CIDR}}. \bibinfo{publisher}{ACM}.
\newblock


\bibitem[\protect\citeauthoryear{Liu, Tong, Qiu, Liu, and Ding}{Liu
  et~al\mbox{.}}{2017}]%
        {liu2017wireless}
\bibfield{author}{\bibinfo{person}{Yu Liu}, \bibinfo{person}{Kin-Fai Tong},
  \bibinfo{person}{Xiangdong Qiu}, \bibinfo{person}{Ying Liu}, {and}
  \bibinfo{person}{Xuyang Ding}.} \bibinfo{year}{2017}\natexlab{}.
\newblock \showarticletitle{Wireless mesh networks in IoT networks}. In
  \bibinfo{booktitle}{\emph{iWEM}}. \bibinfo{publisher}{IEEE}.
\newblock


\bibitem[\protect\citeauthoryear{Mayer, Gupta, Saurez, and Ramachandran}{Mayer
  et~al\mbox{.}}{2018}]%
        {fogstore}
\bibfield{author}{\bibinfo{person}{Ruben Mayer}, \bibinfo{person}{Harshit
  Gupta}, \bibinfo{person}{Enrique Saurez}, {and} \bibinfo{person}{Umakishore
  Ramachandran}.} \bibinfo{year}{2018}\natexlab{}.
\newblock \showarticletitle{{FogStore: Toward a distributed data store for Fog
  computing}}.
\newblock \bibinfo{journal}{\emph{2017 IEEE Fog World Congress, FWC 2017}}
  (\bibinfo{date}{5} \bibinfo{year}{2018}), \bibinfo{pages}{1--6}.
\newblock
\urldef\tempurl%
\url{https://doi.org/10.1109/FWC.2017.8368524}
\showDOI{\tempurl}


\bibitem[\protect\citeauthoryear{Moysiadis, Sarigiannidis, and
  Moscholios}{Moysiadis et~al\mbox{.}}{2018}]%
        {Moysiadis2018TowardsComputing}
\bibfield{author}{\bibinfo{person}{Vasileios Moysiadis},
  \bibinfo{person}{Panagiotis Sarigiannidis}, {and} \bibinfo{person}{Ioannis
  Moscholios}.} \bibinfo{year}{2018}\natexlab{}.
\newblock \bibinfo{title}{{Towards Distributed Data Management in Fog
  Computing}}.
\newblock
\newblock
\showISSN{15308677}


\bibitem[\protect\citeauthoryear{Munir, Kansakar, and Khan}{Munir
  et~al\mbox{.}}{2017}]%
        {IFCIoTintegratedFogCloudIoT}
\bibfield{author}{\bibinfo{person}{Arslan Munir}, \bibinfo{person}{Prasanna
  Kansakar}, {and} \bibinfo{person}{Samee~U. Khan}.}
  \bibinfo{year}{2017}\natexlab{}.
\newblock \showarticletitle{{IFCIoT: Integrated Fog Cloud IoT: A novel
  architectural paradigm for the future Internet of Things}}.
\newblock \bibinfo{journal}{\emph{IEEE Consumer Electronics Magazine}}
  \bibinfo{volume}{6}, \bibinfo{number}{3} (\bibinfo{year}{2017}).
\newblock


\bibitem[\protect\citeauthoryear{Nastic, Rausch, Scekic, Dustdar, Gusev,
  Koteska, Kostoska, Jakimovski, Ristov, and Prodan}{Nastic
  et~al\mbox{.}}{2017}]%
        {nastic2017serverless}
\bibfield{author}{\bibinfo{person}{Stefan Nastic}, \bibinfo{person}{Thomas
  Rausch}, \bibinfo{person}{Ognjen Scekic}, \bibinfo{person}{Schahram Dustdar},
  \bibinfo{person}{Marjan Gusev}, \bibinfo{person}{Bojana Koteska},
  \bibinfo{person}{Magdalena Kostoska}, \bibinfo{person}{Boro Jakimovski},
  \bibinfo{person}{Sasko Ristov}, {and} \bibinfo{person}{Radu Prodan}.}
  \bibinfo{year}{2017}\natexlab{}.
\newblock \showarticletitle{A serverless real-time data analytics platform for
  edge computing}.
\newblock \bibinfo{journal}{\emph{IEEE Internet Computing}}
  \bibinfo{volume}{21}, \bibinfo{number}{4} (\bibinfo{year}{2017}).
\newblock


\bibitem[\protect\citeauthoryear{Pfandzelter and Bermbach}{Pfandzelter and
  Bermbach}{2020}]%
        {tinyFaas}
\bibfield{author}{\bibinfo{person}{Tobias Pfandzelter} {and}
  \bibinfo{person}{David Bermbach}.} \bibinfo{year}{2020}\natexlab{}.
\newblock \showarticletitle{{TinyFaaS: A Lightweight FaaS Platform for Edge
  Environments}}. In \bibinfo{booktitle}{\emph{ICFC}}.
  \bibinfo{publisher}{IEEE}.
\newblock


\bibitem[\protect\citeauthoryear{Plebani, Garcia-Perez, Anderson, Bermbach,
  Cappiello, Kat, Pallas, Pernici, Tai, and Vitali}{Plebani
  et~al\mbox{.}}{2017}]%
        {DitasApproach}
\bibfield{author}{\bibinfo{person}{Pierluigi Plebani}, \bibinfo{person}{David
  Garcia-Perez}, \bibinfo{person}{Maya Anderson}, \bibinfo{person}{David
  Bermbach}, \bibinfo{person}{Cinzia Cappiello}, \bibinfo{person}{Ronen~I.
  Kat}, \bibinfo{person}{Frank Pallas}, \bibinfo{person}{Barbara Pernici},
  \bibinfo{person}{Stefan Tai}, {and} \bibinfo{person}{Monica Vitali}.}
  \bibinfo{year}{2017}\natexlab{}.
\newblock \showarticletitle{{Information logistics and fog computing: The DITAS
  approach}}. In \bibinfo{booktitle}{\emph{CAiSE}}.
  \bibinfo{publisher}{Springer}.
\newblock


\bibitem[\protect\citeauthoryear{Rausch, Rashed, and Dustdar}{Rausch
  et~al\mbox{.}}{2021}]%
        {rausch2021optimized}
\bibfield{author}{\bibinfo{person}{Thomas Rausch}, \bibinfo{person}{Alexander
  Rashed}, {and} \bibinfo{person}{Schahram Dustdar}.}
  \bibinfo{year}{2021}\natexlab{}.
\newblock \showarticletitle{Optimized container scheduling for data-intensive
  serverless edge computing}.
\newblock \bibinfo{journal}{\emph{Future Generation Computer Systems}}
  \bibinfo{volume}{114} (\bibinfo{year}{2021}).
\newblock


\bibitem[\protect\citeauthoryear{Sheikh~Sofla, Haghi~Kashani, Mahdipour, and
  Faghih~Mirzaee}{Sheikh~Sofla et~al\mbox{.}}{2021}]%
        {SheikhSofla2021TowardsComputing}
\bibfield{author}{\bibinfo{person}{Maryam Sheikh~Sofla},
  \bibinfo{person}{Mostafa Haghi~Kashani}, \bibinfo{person}{Ebrahim Mahdipour},
  {and} \bibinfo{person}{Reza Faghih~Mirzaee}.}
  \bibinfo{year}{2021}\natexlab{}.
\newblock \showarticletitle{{Towards effective offloading mechanisms in fog
  computing}}.
\newblock \bibinfo{journal}{\emph{Multimedia Tools and Applications}}
  (\bibinfo{year}{2021}).
\newblock


\bibitem[\protect\citeauthoryear{Shi and Dustdar}{Shi and Dustdar}{2016}]%
        {shi2016promise}
\bibfield{author}{\bibinfo{person}{Weisong Shi} {and} \bibinfo{person}{Schahram
  Dustdar}.} \bibinfo{year}{2016}\natexlab{}.
\newblock \showarticletitle{The promise of edge computing}.
\newblock \bibinfo{journal}{\emph{Computer}} \bibinfo{volume}{49},
  \bibinfo{number}{5} (\bibinfo{year}{2016}).
\newblock


\bibitem[\protect\citeauthoryear{Steffenel}{Steffenel}{2019}]%
        {p2pEdgeDataLocality}
\bibfield{author}{\bibinfo{person}{Luiz~Angelo Steffenel}.}
  \bibinfo{year}{2019}\natexlab{}.
\newblock \showarticletitle{{Improving the Performance of Fog Computing Through
  the Use of Data Locality}}. In \bibinfo{booktitle}{\emph{SBAC-PAD}}.
  \bibinfo{publisher}{IEEE}.
\newblock


\bibitem[\protect\citeauthoryear{Vaquero and Rodero-Merino}{Vaquero and
  Rodero-Merino}{2014}]%
        {vaquero2014finding}
\bibfield{author}{\bibinfo{person}{Luis~M Vaquero} {and} \bibinfo{person}{Luis
  Rodero-Merino}.} \bibinfo{year}{2014}\natexlab{}.
\newblock \showarticletitle{Finding your way in the fog: Towards a
  comprehensive definition of fog computing}.
\newblock \bibinfo{journal}{\emph{ACM SIGCOMM computer communication Review}}
  \bibinfo{volume}{44}, \bibinfo{number}{5} (\bibinfo{year}{2014}).
\newblock


\bibitem[\protect\citeauthoryear{Wang, Zhou, Liu, Bhuiyan, Wang, and Jia}{Wang
  et~al\mbox{.}}{2019}]%
        {Wang2019Fog-basedIoT}
\bibfield{author}{\bibinfo{person}{Tian Wang}, \bibinfo{person}{Jiyuan Zhou},
  \bibinfo{person}{Anfeng Liu}, \bibinfo{person}{Md~Zakirul~Alam Bhuiyan},
  \bibinfo{person}{Guojun Wang}, {and} \bibinfo{person}{Weijia Jia}.}
  \bibinfo{year}{2019}\natexlab{}.
\newblock \showarticletitle{{Fog-based computing and storage offloading for
  data synchronization in IoT}}.
\newblock \bibinfo{journal}{\emph{IEEE Internet of Things Journal}}
  \bibinfo{volume}{6}, \bibinfo{number}{3} (\bibinfo{year}{2019}).
\newblock


\bibitem[\protect\citeauthoryear{Zeuch, Chaudhary, Del~Monte, Gavriilidis,
  Giouroukis, Grulich, Bre{\ss}, Traub, and Markl}{Zeuch et~al\mbox{.}}{2019}]%
        {zeuch2019nebulastream}
\bibfield{author}{\bibinfo{person}{Steffen Zeuch}, \bibinfo{person}{Ankit
  Chaudhary}, \bibinfo{person}{Bonaventura Del~Monte},
  \bibinfo{person}{Haralampos Gavriilidis}, \bibinfo{person}{Dimitrios
  Giouroukis}, \bibinfo{person}{Philipp~M Grulich}, \bibinfo{person}{Sebastian
  Bre{\ss}}, \bibinfo{person}{Jonas Traub}, {and} \bibinfo{person}{Volker
  Markl}.} \bibinfo{year}{2019}\natexlab{}.
\newblock \showarticletitle{The nebulastream platform: Data and application
  management for the internet of things}.
\newblock \bibinfo{journal}{\emph{arXiv preprint arXiv:1910.07867}}
  (\bibinfo{year}{2019}).
\newblock


\end{thebibliography}

\end{document}